\documentclass[10pt, conference, compsocconf]{IEEEtran}

\usepackage[utf8]{inputenc}
\usepackage{amsmath,amssymb}

\usepackage{graphicx}
\usepackage{subfigure}
\usepackage{color}
\usepackage{array}

\usepackage{algorithm}
\usepackage{algorithmic}

\begin{document}

\title{LUNES: Agent-based Simulation of P2P Systems\\(Extended Version)\footnotemark}

\author{\IEEEauthorblockN{Gabriele D'Angelo, Stefano Ferretti}
\IEEEauthorblockA{Department of Computer Science, University of Bologna\\
Bologna, Italy\\
g.dangelo@unibo.it, sferrett@cs.unibo.it}
}

\maketitle

\footnotetext{The publisher version of this paper is available at \url{http://dx.doi.org/10.1109/HPCSim.2011.5999879}. \textbf{{\color{red}Please cite as: Gabriele D'Angelo, Stefano Ferretti. LUNES: Agent-based Simulation of P2P Systems. Proceedings of the International Workshop on Modeling and Simulation of Peer-to-Peer Architectures and Systems (MOSPAS 2011). As part of The 2011 International Conference on High Performance Computing and Simulation (HPCS 2011), ISBN 978-1-61284-382-7.}}}

\begin{abstract}
We present LUNES, an agent-based Large Unstructured NEtwork Simulator, which allows to simulate complex networks composed of a high number of nodes. LUNES is modular, since it splits the three phases of network topology creation, protocol simulation and performance evaluation. This permits to easily integrate external software tools into the main software architecture. The simulation of the interaction protocols among network nodes is performed via a simulation middleware that supports both the sequential and the parallel/distributed simulation approaches. In the latter case, a specific mechanism for the communication overhead-reduction is used; this guarantees high levels of performance and scalability. To demonstrate the efficiency of LUNES, we test the simulator with gossip protocols executed on top of networks (representing peer-to-peer overlays), generated with different topologies. Results demonstrate the effectiveness of the proposed approach. 
\end{abstract}


\section{Introduction}
\label{sec:introduction}

One of the most interesting recent trends in computer science is concerned with the issue of modeling and understanding complex networks. A main reason is that complex connectivity patterns have been observed in most real biological and technological networks~\cite{Barabasi2000,simutools,Newman03thestructure}.
These networks may represent different types of contacts among different entities; for instance, edges in a network can represent sexual contacts, hyper-textual links in Web pages, communication links in peer-to-peer architectures, links among routers in the Internet, ad-hoc connections in wireless sensor networks. All these networks are usually composed of a large amount of nodes and thus, tools that manage the representation of these networks must be able to deal with such large numbers. This does not represent a big issue from a mathematical point of view; complex networks models are in fact built by considering an infinite number of nodes. The problem is more evident when these networks are to be simulated, together with some interaction mechanism among network nodes~\cite{conf/nca/GarbinatoRT07,verma}.\\

In this case, the amount of memory necessary to model such networks is often huge such as the communication requirements among the nodes. This means that also the topology of the network influences the ability to simulate them. Take for instance the case of scale-free networks, i.e.~those nets whose distribution of the node degrees (number of neighbors of a given node) follows a power law function. In this case, the problem of simulating large networks is particularly striking. In fact, the presence of a non-negligible number of nodes with a high degree in a scale-free network, and in general the heterogeneity of nodes, is a key issue that may imbalance the computation and communication load. This may lead to very unsatisfactory results also when parallel or distributed simulation is employed.\\

In a previous work, we have presented PaScaS (Parallel and distributed Scale-free Network Simulator), a simulator able to represent and manage large scale-free networks. That simulator was based on a fast method to build scale-free networks and to model information sharing and application protocols above them. PaScaS was built on top of a simulation middleware for the implementation of both sequential (i.e.~monolithic) and Parallel And Distributed Simulation (PADS)~\cite{simutools}.\\

Following the experience of PaScaS, in this work we present an agent-based discrete-event simulator, called LUNES (Large Unstructured NEtwork Simulator). The main goal of LUNES is to offer an efficient and easy-to-use tool for the simulation of complex protocols on top of large graphs of whatever topology (not only scale-free nets). In practice, LUNES is able to import the graph topologies generated by other tools (e.g.~igraph) and provides the functionalities that are needed for the performance evaluation of simulated protocols. One of the main goals of the simulator design is to obtain a tool that clearly splits the fundamental phases:
\begin{itemize}
\item network topology creation;
\item protocol simulation in a specific testbed;
\item traces analysis (i.e.~performance evaluation).
\end{itemize}

This modular approach permits the easy integration of external software tools. In practice, such integration is based on very simple template files (such as the graphviz dot language~\cite{graphviz}) and provides a good level of extensibility. Under the performance and scalability viewpoint, the most demanding points are the protocol simulation and the traces analysis. The traces analysis has been excluded from the simulation tasks and specific software tools have been implemented for this purpose. All such tools have been designed to work in parallel, exploiting all the computational resources provided by parallel (multi-processor or multi-core) architectures. Under the implementation viewpoint, LUNES is a complete re-engineering and re-implementation since it does not share any software component with PaScaS.\\

In LUNES, the simulation services are provided by the ART\`IS middleware and the GAIA framework~\cite{gdapads03,artis}. In this way, the LUNES user does not need to deal with low-level simulation details and can take advantage of all the features offered by the middleware and the framework. More in detail, the ART\`IS middleware is in charge of providing all the necessary services to build up sequential and parallel/distributed simulations (e.g.~synchronization). The role of GAIA is to provide some advanced features such as the dynamic model partitioning and the load-balancing; all this is transparent to the simulator user.\\

In this paper, we evaluate the tool by simulating unstructured peer-to-peer overlays as complex networks, and by running on top of them data dissemination schemes based on gossip protocols. Results show the effectiveness of the proposed approach. It is worth noting that, LUNES in both binary and source code versions can be downloaded from the project website~\cite{artis}. As usual, we aim to provide all the tools, configurations and scenarios used to conduct the performance evaluation shown in this paper. The version currently provided is still an early preview; in the next months we plan to include better documentation and, more importantly, to include more user level protocols and network construction methods.\\

The remainder of this paper is organized as follows. In Section~\ref{sec:background} we present some background related to complex networks, their modeling and simulation. We also discuss our previous work on the field. Section~\ref{sec:lunes} describes our novel simulator, LUNES. In Section~\ref{sec:exp} we report on an evaluation we performed to assess the performances of LUNES. Finally, in Section~\ref{sec:conc} we provide some final remarks.

\section{Background}
\label{sec:background}

\subsection{Simulation of Complex Networks}

The theory of complex networks has been broadly investigated in literature~\cite{Barabasi2000,simutools,Newman03thestructure}. Despite these strong efforts on the characterization and modeling of these nets, there are only few works on specific simulators of complex networks. Our goal is to provide a scalable and easy-to-use simulator that can be used to design new interaction protocols among nodes of a network.\\

We already mentioned our previous work, specifically designed for the construction of scale free networks~\cite{simutools}. We describe this tool and its main features in Section~\ref{sec:pascas}. Another paper worth of mention is \cite{KincaidA05}, which shows how to build a scale-free network to simulate air transportation networks. Focus is given on the scheme to build the network, rather than the need to have an effective simulation tool itself. Instead, \cite{dobrescu} proposes a model to simulate scale-free networks; they start from a simulator running on a single-CPU, and compare it with a distributed environment with parallel clustered processors.\\

Turning the focus on the simulation of interaction protocol on communication networks, an existing software for the simulation of peer-to-peer overlays is Peersim \cite{p2p09-peersim}. This tool can be employed to mimic gossip algorithms and other communication protocols only. In this sense, the purposes of the tool are similar to those of LUNES; however, the software architecture is very different, since it is based on a single running process that mimics the behavior of several agents.  On the field, other solutions related to agent-based simulation have been presented in~\cite{Niazi:2009,Takahashi:2006}.

\subsection{Parallel and Distributed Simulation}

The task of simulating huge, complex networks poses many challenges. The amount of computation and memory that is necessary to model such systems is so high that traditional simulation approached based on the sequential (i.e.~monolithic) execution of simulation models are unable to provide the necessary scalability. In many other cases, this issue has been faced by resorting to Parallel and Distributed Simulation (PADS)~\cite{FUJ00}. Following the PADS approach, the simulation model is partitioned among a set of execution units (i.e.~CPUs). Obviously, to obtain meaningful results all the units that are involved in the simulation have to be synchronized such as all the interactions among parts of the simulated model. The main advantage of PADS is that, with respect to monolithic, much more resources can be used for modeling the system to simulate and to complete the simulation as fast as possible. On the other side, this approach is not free from drawbacks: the main one is the communication cost due to 
interactions among simulation parts that now are allocated on different CPUs or execution nodes (e.g.~in a LAN-based execution architecture). Under the performance viewpoint, what often happens is that the cost for maintaining synchronized the parallel/distributed execution architecture and to provide the updates to the whole simulation model is so high that the distributed simulator has poor performance. In other words, it is a problem that is very hard to be parallelized. It happens quite often that, given a specific system to be simulated, the sequential implementation of the simulator results faster than the PADS version.\\

Many works have addressed the problem of computation load-balancing in parallel simulation environments~\cite{deelman98,boukerche97,gan00,shanaker01,som00} but only a few have considered both communication cost and load-balancing requirements in PADS environments. For example, in \cite{peschlow07} a dynamic partitioning algorithm is proposed for optimistic distributed simulation. At present time, the most parallel and distributed simulators employ a static partitioning of the simulation models. That mean that such tools are unable to dynamically adapt to the simulation model behavior and to react to imbalances in the model and in the execution architecture. In other words, it is necessary to predict the behavior of each part of the model and to execute the simulation in a strictly controlled environment.\\

In the last years, our research group has designed and implemented a new PADS-based software architecture. The Advanced RTI System (ART\`IS) is a middleware that can be used to build sequential simulations but with some specific features aimed to build efficient parallel and distributed simulations. Its design is partially inspired by the High Level Architecture (HLA, IEEE 1516 standard)~\cite{ieee1516}, but in practice it has been used as a testbed for implementing and evaluating new features that are missing from the standard.\\

On top of the ART\`IS middleware there is the Generic Adaptive Interaction Architecture (GAIA)~\cite{gdapads03} framework. It provides to the simulation developer a template and some services to build models following the discrete event simulation approach. The implemented paradigm is agent based: the simulation model is partitioned in a set of Simulated Model Entities (SME) and each of them is in practice an agent that interacts with other agents using timestamped messages. Other than such aspects, the main characteristics of GAIA is the implementation of a migration based environment that is able to cluster (migrate) the highly interacting entities (i.e.~agents) in the same execution units, and therefore reduce the communication overhead.  Since this aspect is of fundamental importance in LUNES, in the next subsection some further details about the clustering mechanism will be provided.

\subsection{The Clustering Mechanism}

The core of the clustering mechanism is the partitioning of the simulated model in a set of parts (e.g.~the SMEs). Each SME models the evolution of a portion of the system and interacts with other SMEs following a message passing based approach. Basically, the goal of the mechanism, when applied in a parallel or distributed environment, is to cluster the highly interacting SMEs in the same execution unit, so as to reduce the communication cost. What happens is that the SMEs allocated in the same CPU will be able to communicate with low latency and high bandwidth. Conversely, all the interactions that will involve more than one CPU or different hosts will experience a higher communication overhead. The clustering mechanism is in charge of adaptively reallocating (i.e. migrating) the SMEs over the available execution units. In this way, it is possible to reduce the overall cost of communication with the effect of reducing the amount of time that is needed for completing the simulation runs (i.e.~Wall-Clock-
Time, WCT).\\

Given that the behavior of each SME is not predictable a priori, the clustering mechanism is based on the auditing of the communication pattern of each SME during the simulation execution. A set of specific heuristics evaluates if reallocations are necessary and if they will provide a benefit in terms of communication overhead. It is worth noting that, each migration of SMEs has a cost that is due to many factors such as the state variables of the SME and the amount of time that is needed to complete the data transfer. All the migration procedures have to be executed without altering in any way the semantics of the simulation model and with the least possible effort from the simulation modeler.\\

Up to now, we have considered only the communication overhead and how to manage its reduction; another main aspect worth of consideration is the computation load-balancing. The idea is that the clustering described above is constrained by the load-balancing requirements of the parallel or distributed execution architecture. In other words, the clustering of SMEs must take care of the load of each execution unit involved in the PADS. In practice, it is not possible to cluster all the interacting SMEs together and no part of the execution architecture can be overloaded.\\

All the mechanisms described above have to be implemented considering that: i) the number of SMEs could be very large; ii) the clustering heuristics have to be so general to work with a wide range of simulation models and without any knowledge of the model semantic. For these reasons, in our experience very simple solutions have to be preferred to more complex ones. More details about GAIA, its features and implementation can be found in \cite{gdadsrt04} and \cite{gdaijspm}.

\subsection{PaScaS}
\label{sec:pascas}

PaScaS (Parallel and distributed Scale-free Network Simulator) is a tool specifically designed for the modeling of scale-free networks~\cite{simutools}. The graph generation is obtained using the algorithm proposed by Bara\-b\'a\-si and Albert in \cite{barab99}. Based on ART\`IS, PaScaS can be used to implement both sequential and parallel/distributed simulations. The core of PaScaS is a simulation model implemented (in C language) using the APIs provided by GAIA. The model implements the main features of the scale-free network simulator such as the building algorithm, the behavior and the characteristics of each node and the gossiping protocols.\\

The set up of the simulated scenarios, and the tuning of the runtime parameters of the simulator, is obtained via configuration files and environment variables. A set of scripts is provided to facilitate and automatize the execution of parallel and distributed runs. This approach has been chosen to facilitate the set up of unattended batch executions. The results of the runs are collected in logging files, tuned to the adequate level of detail that has been chosen by the simulation modeler. Other scripts are available to collect and analyze the requested data and results.\\

Under the performance viewpoint, PaScaS has shown quite interesting results~\cite{simutools} but its usability and extensibility was limited by some design and implementation choices. The release of a new version of the GAIA framework, providing a better and easier environment for the implementation of migration-enabled simulation modes, has fostered a complete redesign and re-implementation of the simulator. This effort has led to the creation of LUNES that will be introduced in the following section.

\section{LUNES}
\label{sec:lunes}

LUNES (Large Unstructured NEtwork Simulator) is an easy-to-use tool for the simulation of complex protocols on top of large graphs of whatever topology. It is modular and separates into different software components the tasks of: i) network creation, ii) implementation of the protocols and iii) analysis of the results. The use of a modular approach has the advantage of permitting the (re-)use and integration of existing software tools, and facilitates the update and extensibility of the tool. The flow of data processing is linear, i.e.~a network is created by the network creation topology module; then a communication protocol is executed on top of such a network by the protocol simulator; its results are analyzed by the trace analysis module. It is worth mentioning that all such tools have been designed and implemented to work in parallel and therefore are able to exploit all the computational resources provided by parallel (multi-processor or multi-core) or distributed (e.g.~clusters of PCs) architectures. 
In other words, while for instance a network (generated by the network creation topology module) is exploited by the protocol simulator, the network creation topology module may be active for the generation of another network. Similarly, while the protocol simulator is running, its outcomes from previous executions can be analyzed by the trace analysis module. Outcomes from a given module are exploited by the other one via simple template files (such as the graphviz dot language~\cite{graphviz}). These modules are described in isolation in the rest of the section.

\subsection{Network Topology Creation}

LUNES is able to import the graph topologies generated by other tools. In the current version of LUNES, we employ \textit{igraph}, an interesting tool for creating and manipulating undirected and directed graphs~\cite{igraph}. It includes algorithms for network analysis methods and graph theory and allows to handle graphs with millions of vertices and edges. The graphs generated by igraph or other tools can be directly used for protocol simulation or much more often are stored in ``corpuses''. Each corpus can be seen as a testbed environment in which compare the behavior and outcomes of protocols under exactly the same conditions. Using an external tool for the generation of graph topologies does not mean that the analyzed graphs have to be static. During the simulation execution it is always possible to modify the network topology and to deal with dynamic systems.

\subsection{Protocol Simulation} 
\label{sec:protocol_simulation}

As said above, in LUNES, the simulation services are demanded to ART\`IS and GAIA. This means that, in the implementation of new protocols to be simulated in LUNES, there is no need for dealing with low-level simulation details. The only Application Programming Interface (API) used in LUNES is quite high level and is provided by GAIA. Furthermore, for the implementation of new protocols LUNES already offers a set of primitives and functions that can be used and modified without the need of starting from scratch. For example, in the current version all the most common features of dissemination protocols are already implemented and adding new variants or protocols is straightforward.

\subsection{Trace Analysis}

Under the performance and scalability viewpoint, the most demanding points are the protocol simulation and the traces analysis. As to the traces analysis, it has been excluded from the simulation tasks and some specific software tools have been implemented. The simulation of a network with a few hundred nodes for the time necessary for studying some common properties can generate a huge amount of simulation traces that have to be stored, parsed and analyzed (in the order of few gigabytes per run). This means that, very simple metrics used for performance evaluation of the simulated protocol can require a lot of effort. In the current version of LUNES, this task is implemented using a set of shell scripts and some specific tools that have been implemented in C language for efficiency. This mix is both quite efficient and easy to extend and personalize. We have intentionally avoided to build a monolithic application to provide users with an easily customizable tool.

\section{Performance Evaluation}
\label{sec:exp}

In this section we investigate the performance of LUNES by running on top of it some gossip-based communication protocols for unstructured peer-to-peer architectures. It is well-known that gossip schemes can easily spread information through networks, and are widely studied in the context of unstructured peer-to-peer networks \cite{simutools,disio2011,disio}. In essence, according to this communication paradigm, all nodes are able to generate a new message to be disseminated in the network. When the generation procedure is invoked at a given node, a single message is created, with a certain probability. The generation of a message triggers the occurrence of a new event produced at a given node that must be propagated. If the message is created, then it is gossiped through the net, using a \textsc{gossip()} procedure.\\

We now review the two different gossip schemes, employed to test our simulation architecture; more details on these schemes can be found in \cite{disio2011,disio}.

\subsection{The Communication Protocols}

The first dissemination protocol we employ is referred as \emph{Fixed Probability}. It is very simple: as soon as a node (say $p$) needs to propagate a message \emph{msg}, all $p$'s neighbors are considered and a threshold value $\mathit{v} \leq 1$ is maintained, which determines the probability that \emph{msg} is gossiped to the neighbor~\cite{simutools,conf/nca/GarbinatoRT07,verma}.\\

The second gossip scheme is referred as \emph{Adaptive Gossip}~\cite{disio2011}. In essence, once a peer receives a message from a neighbor, it forwards the message to all other neighbors (i.e.~a broadcast is performed) based on a dissemination probability. However, as soon as a node $p$ observes that it is receiving messages from another peer $q$ at a rate lower than expected, it activates a countermeasure, asking its neighborhood (actually, the neighbor $n$ from which it usually receives messages originated from $q$, or a random neighbor if it did not receive any message at all) to increase its dissemination probability of game events coming from $q$ and that will be delivered to $p$. The request from $p$ to $n$ to increase the dissemination probability can be interpreted as a stimulus that remains active at $n$ for a limited period of time. Then, the dissemination probability returns to the original value (i.e.~the stimulus decades in time). This approach is adopted to avoid that in time all dissemination 
probabilities reach the maximum value and thus the gossip scheme becomes a pure broadcast algorithm.\\

From a simulation point of view, it is worth noting that the employed simulative scenarios require a high communication among nodes in the simulated network. Hence, they represent important use cases to benchmark LUNES. As to the computation, the \emph{Fixed Probability} scheme requires a low computation, since the nodes' behavior is quite simple, while the \emph{Adaptive Gossip} is more computation demanding.

\subsection{Evaluation and Results}

\begin{figure*}[htp]
  \begin{center}
    \subfigure[Fixed probability dissemination]{\label{fig:fixed_scalability}\includegraphics[angle=0,width=0.48\linewidth]{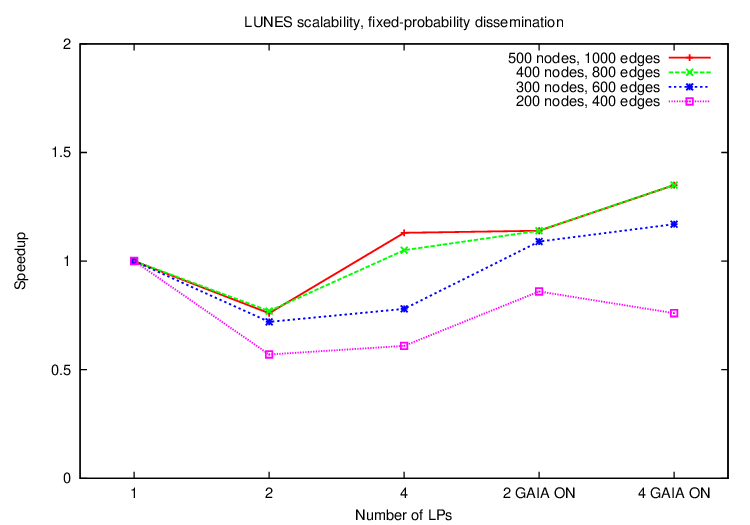}}
    \subfigure[Adaptive gossip dissemination]{\label{fig:adaptive_scalability}\includegraphics[angle=0,width=0.48\linewidth]{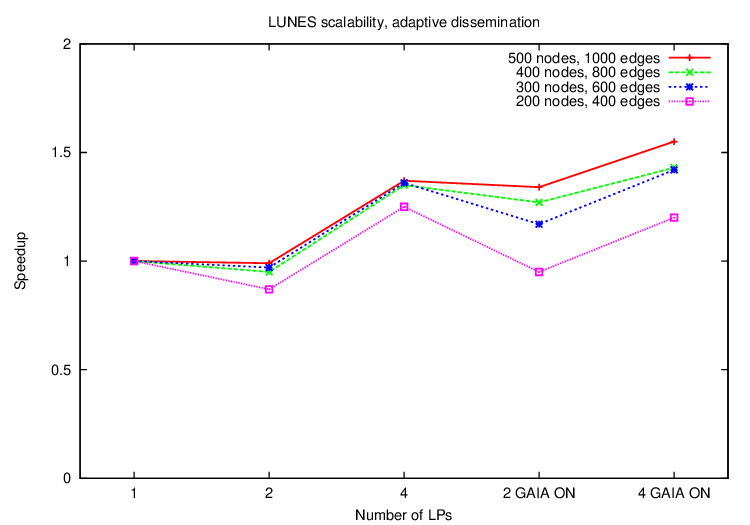}}
  \end{center}
  \caption{Scalability of the LUNES simulator with different dissemination protocols.}
  \label{fig:scalability}
\end{figure*}

The first step in the performance evaluation of LUNES has been the creation of a set of scenarios that will be used as testbed. As illustrated in Table~\ref{table:wct}, 4 different scenarios have been generated. In practice, each scenario is composed of 10 graphs (a corpus) with common characteristics in terms of number of nodes ($n$) and edges ($e$). For this performance evaluation, it has been chosen to build all the graphs using the random (Erdos-Renyi game) graph generator provided in the igraph software. Each measure shown in Table~\ref{table:wct} and Figure~\ref{fig:scalability} refers to the evaluation of the whole corpus and not to a single graph. The different communication protocols described previously have been run on each testbed and the amount of simulated time has been set to 1000 timesteps. The TTL (Time-To-Live) that is in Table~\ref{table:wct} refers to the presence in each evaluated communication protocol of a counter that limits the number of hops that a message is permitted to travel 
before being discarded by a host. This value has a big impact on both the communication protocols and the simulator performance. In fact, it increases the number of messages that are sent on the network; this means that the higher the TTL the more difficult is to achieve good performance in a PADS. In the scenarios of Table~\ref{table:wct}, the TTL value has been always chosen as an approximate value of the diameter of the network, given by $\lceil \frac{\ln n}{\ln \lambda} \rceil$, being $n$ the number of nodes in the network and $\lambda$ the mean degree of network nodes~\cite{Chung:2001}.\\

\begin{figure}[t]
\centering
 \includegraphics[angle=-90,width=\linewidth]{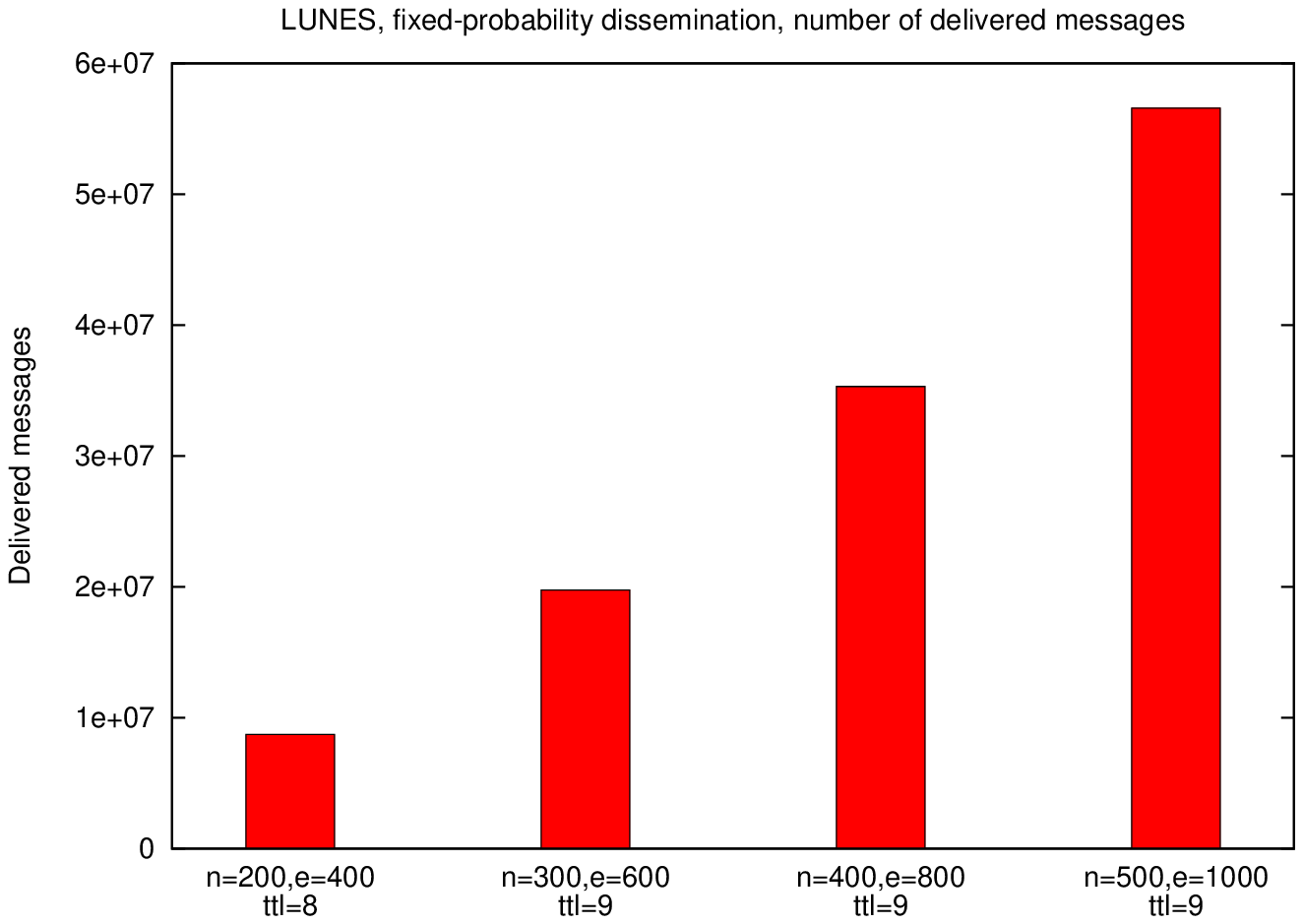}
\caption{Number of delivered messages in the analyzed scenarios.}
\label{fig:fixed_probability_messages}
\end{figure}

In all cases presented in this section, the baseline dissemination probability for both the Fixed Probability and the Adaptive Gossip protocols was set to $0.8$. This means that each message received by a node has (at least) a 80\% probability to be forwarded to each of the neighbors of such node. Under the simulator viewpoint this means a huge number of messages to be created, delivered and processed. It is worth noting that, as shown in Figure~\ref{fig:fixed_probability_messages} the number of delivered message increases sharply when adding more nodes and edges to the network. Doubling the number of nodes has the effect of generating a number of messages that is more than double, and this effect is further increased when, due to an increment of the network size, a higher TTL is set at nodes. In practice, all this means that an increment of the network size has the effect of augmenting significantly the amount of communication that the PADS will have to cope with, in a way such that communication overwhelms 
the computation requirements. Larger graphs are harder and harder to simulate in a parallel or distributed simulation environment.\\

The results shown in the following of this section have been collected using an Intel(R) Xeon(R) CPU X3220 (2.40GHz) with 4 cores, 4 GB of RAM and equipped with Ubuntu 8.04.4 LTS (x86\_64 GNU/Linux, version 2.6.24-27-server SMP). Each measurement refers only to the protocol simulation (as described in subsection~\ref{sec:protocol_simulation}). In fact, including also the network topology creation and the trace analysis would have been useless, given that both are much easier to parallelize with respect to the protocol simulation. Under the statistical viewpoint, we performed multiple runs for each experiment, and the confidence intervals obtained with a 95\% confidence level are lower than 5\% the average value of the performance index shown.\\

\begin{table*}[t]
\caption{Wall Clock Time (seconds), sequential simulation (LP=1)}
\centering
\begin{tabular}{c|c|c|c|c|c}
\hline
\textbf{Scenario} & \textbf{\# Nodes ($n$)} & \textbf{\# Edges ($e$)} & \textbf{$TTL$} & \textbf{WCT for Fixed Probability (sec)} & \textbf{WCT for Adaptive Gossip (sec)} \\
\hline
1 & $200$ & $400$ & $8$ & $93.326$ & $150.43$  \\
2 & $300$ & $600$ & $8$ & $240.579$ & $357.777$  \\
3 & $400$ & $800$ & $8$ & $480.673$  & $664.348$ \\
4 & $500$ & $1000$ & $9$ & $820.657$ & $1153.895$ \\
\hline
\end{tabular}
\label{table:wct}
\end{table*}

Table~\ref{table:wct} shows the amount of time that is necessary to complete the simulation of each of the scenarios (in each case a corpus of 10 graphs). Both the communication protocols (Fixed Probability and Adaptive Gossip) have been simulated using a monolithic approach. This is usually referred as $LP=1$ and it means that a single Logical Process (LP) is responsible to manage the whole simulation model. In the following of this section, we consider other configurations such as $LP={2,4}$. Each LP is allocated on a single CPU-core and, in practice, this means that we will consider a sequential simulation ($LP=1$), a parallel one using 2 cores ($LP=2$) a finally a configuration configuration with 4 cores ($LP=4$). The results shown in the table confirm that the Fixed Probability is much less demanding in terms of computation with respect to the Adaptive Gossip. In both cases it possible to say that average size graphs can be simulated in a quite short time frame.
As stated above, in the following of this performance evaluation it will be investigated if a PADS approach is able to offer valuable results in a scenario that is so challenging.\\

Figure~\ref{fig:scalability} (a) shows the results obtained by LUNES in the simulation of the Fixed Probability protocol in the sequential ($LP=1$) and parallel configurations ($LP=2,4$) described above. Furthermore, two configurations referred as ``2 GAIA ON'' and ``4 GAIA ON'' are shown; they refer to parallel configurations with the clustering mechanism (GAIA) turned ON. Similar scenarios are reported in Figure~\ref{fig:scalability} (b), that refers to the evaluation of the Adaptive Gossip protocol. In both Figures, the evaluation metric is the speedup, i.e.~how much a parallel/distributed configuration is faster than the corresponding sequential one. In other words, if $speedup=1$ then the parallel implementation has the same performance of the sequential one, if $speedup>1$ then the parallelization improves the simulation performance and, finally, if $speedup<1$ then sequential is better.\\

Figure~\ref{fig:scalability} (a) clearly shows that a parallel configuration with $LP=2$ gives poor results. This is true for all the considered scenarios but it is worse when the number of nodes is limited (e.g. 200 nodes, 400 edges, pink line with empty squares). This means that the amount of computation in this scenario is so limited that there is no need for any form of parallelization and therefore that using a PADS approach is useless. Still focusing on $LP=2$ it is possible to see that more computational intensive scenarios (e.g. 500 nodes, 1000 edges, red line with ticks) have slightly better results but also in this case the parallel configuration is slower than the sequential one. Much better results are obtained by the computational intensive scenarios with $LP=4$. This means that, adding more computational resources balances the cost that is due to the synchronization and communication requirements of the PADS architecture. A final comment is for the ``2 GAIA ON'' and ``4 GAIA ON'' configurations:
 in both cases they are able to obtain much better results with respect to the configuration in which the clustering mechanism is disabled. Also in this case, the best results are achieved for scenarios in which the computational requirements are higher. This is correct given that the clustering mechanism is able to reduce the communication overhead but, in the current version, has no impact on the load sharing among the execution units. As a future work we will design and implement an extension of GAIA that will try to understand (at runtime) if the amount of computation required by the simulation model is more adequate for a sequential simulation or a PADS one.\\

As expected, the performance of LUNES in the simulation of the adaptive gossip protocol are still better (Figure~\ref{fig:scalability} (b)). As described before, this protocol has much higher computational requirements with respect to Fixed Probability. In fact, each node has to check the reception rate of messages from other nodes and decides if it is necessary to send stimuli for increasing the dissemination probability of neighbors. All that gives a much higher amount of computation in the simulator and therefore the PADS can give some benefits. In this situation, the configuration with 2 LPs gives results that are near to the sequential configuration. As predicted, $LP=4$ has good results also when the simulated scenario has a few simulated nodes (e.g. 200 nodes, 400 edges, pink line with empty squares). Again, the best results are obtained with ``4 GAIA ON''. Nevertheless, also ``2 GAIA ON'' gives quite good results; for some configurations the performance of ``2 GAIA ON'' are near to the one obtained 
by ``4 GAIA ON'' also if only 2 CPU-cores are used.\\

In view of the characteristics of the models being simulated (which imply very high communication and low computation) we think that such results are quite interesting. In this performance evaluation we basically simulated unstructured P2P gossip protocols, which by definition are quite simple. Many real-world P2P protocols are much more demanding in terms of computation. We claim that for these schemes, resorting to a PADS approach would lead to further advantages. The sequential configuration handles well with scenarios with limited size and complexity, but in the next years it is expected that many cores CPUs will be available in every desktop PC. This is surely a good reason for considering more complex and specific PADS approaches.

\section{Conclusions}
\label{sec:conc}

In this paper, we have presented LUNES, a novel simulator for complex networks composed of a high number of nodes, and to test interaction protocols among network nodes. An agent-based approach is employed; thus, the simulation is specified by implementing the behavior of each simulated node. LUNES is modular, since it splits the three phases of network topology creation, protocol simulation and performance evaluation. We have employed LUNES to simulate gossip protocols run on peer-to-peer overlays. Results demonstrate the effectiveness of LUNES and some promising results when a parallel/distributed simulation approach is followed. In particular, some advanced features such as the dynamic clustering of the simulation model entities have demonstrated to be able to reduce the communication overhead in the execution architecture, that is one of the main critical issues in PADS.\\

As a future work we plan to further extend the LUNES implementation, in particular adding more communication protocols in the software distribution and testing the simulator in more complex scenarios. In the next versions of the GAIA framework more complex clustering mechanisms for both the communication overhead reduction and the computation load-balancing will be provided. Given its structure, LUNES will benefit from such new features without requiring any modification. In particular, our next goal is to provide a tool that is able to automatically choose the best configuration (e.g.~monolithic vs. parallel/distributed) without any extra effort from the user. In our vision, all such details should be transparent to the simulator user, that is much more interested in the implementation and testing of new communication protocols and P2P systems.\\

The choice of using the igraph library in the network creation topology module is currently under discussion. The library provides a good framework for the creation complex network topologies and many useful tools. Despite of this, in the currently used version (0.5.3) we found some problems in the construction of both random (Erdos-Renyi game) and scale-free (Barabási-Albert model) networks. We plan to investigate more in deep our findings and to implement some fixes for igraph or to implement a topology generator specifically designed for LUNES.


\small{
\bibliographystyle{abbrv}
\bibliography{paper}  

\begin{thebibliography}{10}

\bibitem{igraph}
igraph web site.
\newblock http://igraph.sourceforge.net/, 2011 Feb.

\bibitem{barab99}
A.-L. Barab\'{a}si and R.~Albert.
\newblock Emergence of scaling in random networks.
\newblock {\em Science}, 286:509--512, 1999.

\bibitem{Barabasi2000}
A.-L. Barab{\'a}si, R.~Albert, and H.~Jeong.
\newblock Scale-free characteristics of random networks: the topology of the
  world-wide web.
\newblock {\em Physica A: Statistical Mechanics and its Applications},
  281(1-4):69--77, Jun 2000.

\bibitem{gdadsrt04}
L.~Bononi, M.~Bracuto, G.~D'Angelo, and L.~Donatiello.
\newblock A new adaptive middleware for parallel and distributed simulation of
  dynamically interacting systems.
\newblock In {\em Proc. of the 8th IEEE International Symposium on Distributed
  Simulation and Real-Time Applications}. IEEE, 2004.

\bibitem{gdapads03}
L.~Bononi, {G. D'Angelo}, and L.~Donatiello.
\newblock {HLA-based} adaptive distributed simulation of wireless mobile
  systems.
\newblock In {\em Proc. 17th {ACM/IEEE/SCS} Workshop on Parallel and
  Distributed Simulation}. IEEE Press, 2003.

\bibitem{boukerche97}
A.~Boukerche and S.~Das.
\newblock Dynamic load balancing strategies for conservative parallel
  simulations.
\newblock In {\em PADS '97: Proc. of the eleventh workshop on Parallel and
  distributed simulation}. IEEE, 1997.

\bibitem{Chung:2001}
F.~Chung and L.~Lu.
\newblock The diameter of sparse random graphs.
\newblock {\em Adv. Appl. Math.}, 26:257--279, May 2001.

\bibitem{gdaijspm}
G.~D'Angelo and M.~Bracuto.
\newblock Distributed simulation of large-scale and detailed models.
\newblock {\em International Journal of Simulation and Process Modelling},
  5(2):120--131, 2009.

\bibitem{simutools}
G.~D'Angelo and S.~Ferretti.
\newblock Simulation of scale-free networks.
\newblock In {\em Simutools '09: Proc.~of the 2nd International Conference on
  Simulation Tools and Techniques}, pages 1--10, ICST, Brussels, Belgium, 2009.
  ICST.

\bibitem{disio2011}
G.~D'Angelo, S.~Ferretti, and M.~Marzolla.
\newblock Adaptive event dissemination for peer-to-peer multiplayer online
  games.
\newblock In {\em Proc. of the International Workshop on DIstributed SImulation
  and Online gaming (DISIO 2011)}. ICST, March 2011.

\bibitem{deelman98}
E.~Deelman and B.~Szymanski.
\newblock Dynamic load balancing in parallel discrete event simulation for
  spatially explicit problems.
\newblock {\em SIGSIM Simul. Dig.}, 28(1), 1998.

\bibitem{dobrescu}
R.~Dobrescu, S.~Taralunga, and S.~Mocanu.
\newblock Web traffic simulation with scale-free network models.
\newblock In {\em Proc. of 7th WSEAS International Conference on Applied
  Informatics and Communications}, pages 275--280, 2007.

\bibitem{disio}
S.~Ferretti and G.~D'Angelo.
\newblock Multiplayer online games over scale-free networks: a viable solution?
\newblock In {\em Proc.~of the International Workshop on DIstributed SImulation
  and Online gaming (DISIO 2010)}. ICST, 2010.

\bibitem{FUJ00}
R.~Fujimoto.
\newblock {\em Parallel and Distributed Simulation Systems}.
\newblock {\it Wiley \& Sons}, 2000.

\bibitem{gan00}
B.~Gan, Y.~Low, S.~Jain, S.~Turner, W.~Cai, W.~Hsu, and S.~Huang.
\newblock Load balancing for conservative simulation on shared memory
  multiprocessor systems.
\newblock In {\em PADS '00: Proc. of the fourteenth workshop on Parallel and
  distributed simulation}. IEEE, 2000.

\bibitem{graphviz}
E.~R. Gansner and S.~C. North.
\newblock An open graph visualization system and its applications to software
  engineering.
\newblock {\em Softw. Pract. Exper.}, 30:1203--1233, September 2000.

\bibitem{conf/nca/GarbinatoRT07}
B.~Garbinato, D.~Rochat, and M.~Tomassini.
\newblock Impact of scale-free topologies on gossiping in ad hoc networks.
\newblock In {\em NCA}, pages 269--272. IEEE Computer Society, 2007.

\bibitem{ieee1516}
{IEEE 1516 Standard, Modeling and Simulation ({M\&S}) High Level Architecture
  (HLA)}, 2000.

\bibitem{KincaidA05}
R.~K. Kincaid and N.~M. Alexandrov.
\newblock Scale-free networks: A discrete event simulation approach.
\newblock In {\em International Conference on Computational Science}, pages
  1051--1058, 2005.

\bibitem{p2p09-peersim}
A.~Montresor and M.~Jelasity.
\newblock Peersim: A scalable p2p simulator.
\newblock In {\em Proc. of the 9th International Conference on Peer-to-Peer
  (P2P'09)}, pages 99--100, Seattle, WA, Sept. 2009.

\bibitem{Newman03thestructure}
M.~E.~J. Newman.
\newblock The structure and function of complex networks.
\newblock {\em SIAM Review}, 45:167--256, 2003.

\bibitem{Niazi:2009}
M.~Niazi and A.~Hussain.
\newblock Agent-based tools for modeling and simulation of self- organization
  in peer-to-peer, ad hoc, and other complex networks.
\newblock {\em Comm. Mag.}, 47:166--173, March 2009.

\bibitem{peschlow07}
P.~Peschlow, T.~Honecker, and P.~Martini.
\newblock A flexible dynamic partitioning algorithm for optimistic distributed
  simulation.
\newblock In {\em Proc. of the 21st International Workshop on Principles of
  Advanced and Distributed Simulation}. IEEE, 2007.

\bibitem{shanaker01}
M.~Shanaker, R.~Padman, and W.~Kelton.
\newblock Efficient distributed simulation through dynamic load balancing.
\newblock {\em IIE Transactions}, 33(3), 2001.

\bibitem{som00}
T.~Som and R.~Sargent.
\newblock Model structure and load balancing in optimistic parallel discrete
  event simulation.
\newblock In {\em PADS '00: Proc. of the fourteenth workshop on Parallel and
  distributed simulation}. IEEE, 2000.

\bibitem{Takahashi:2006}
T.~Takahashi and H.~Mizuta.
\newblock Efficient agent-based simulation framework for multi-node
  supercomputers.
\newblock In {\em Proceedings of the 38th conference on Winter simulation}, WSC
  '06, pages 919--925. Winter Simulation Conference, 2006.

\bibitem{artis}
{Unibo Parallel and Distributed Simulation (PADS) Research Group Homepage}.
\newblock http://pads.cs.unibo.it, 2011.

\bibitem{verma}
S.~Verma and W.~T. Ooi.
\newblock Controlling gossip protocol infection pattern using adaptive fanout.
\newblock In {\em ICDCS '05: Proceedings of the 25th IEEE International
  Conference on Distributed Computing Systems}, pages 665--674, Washington, DC,
  USA, 2005. IEEE Computer Society.

\end{thebibliography}
}


\end{document}